\documentclass[
	article,			
	12pt,				
	oneside,			
	a4paper,			
	english,			
	brazil,				
	sumario=tradicional
	]{abntex2}

\usepackage{lmodern}			
\usepackage[T1]{fontenc}		
\usepackage[utf8]{inputenc}		
\usepackage{indentfirst}		
\usepackage{nomencl} 			
\usepackage{color}	
\usepackage{url}			
\usepackage{graphicx}			
\usepackage{microtype} 			
		
\usepackage[num]{abntex2cite}	
\newcommand\Rey{\mbox{\textit{Re}}}  

\titulo{T\'uneis de vento num\'ericos\\ Numerical wind tunnels}
\autor{
\hspace{-92pt}Paulo Victor Santos Souza\thanks{paulo.victor@ifrj.edu.br} \\
\hspace{-92pt}Instituto de Ciências Exatas, Universidade Federal Fluminense, Volta Redonda\\
\hspace{-92pt}Instituto Federal de Educação, Ciência e Tecnologia do Rio de Janeiro\\
 \hspace{-92pt} Volta Redonda, Rio de Janeiro, Brasil\\
\hspace{-92pt}Paulo Murilo Castro de Oliveira\thanks{pmco@if.uff.br}\\
\hspace{-92pt}Instituto de Física, Universidade Federal Fluminense\\
\hspace{-92pt} Niterói, Rio de Janeiro, Brasil\\
\hspace{-92pt}Instituto Mercosul de Estudos Avançados, Universidade Federal da Integração Latino Americana\\
\hspace{-92pt} Foz do Iguaçu, Paraná, Brasil
}
\local{Brasil}
\data{06/12/2015}

\definecolor{blue}{RGB}{41,5,195}

\makeatletter
\hypersetup{
		pdftitle={\@title}, 
		pdfauthor={\@author},
    	pdfsubject={Modelo de artigo científico com abnTeX2},
	    pdfcreator={LaTeX with abnTeX2},
		pdfkeywords={abnt}{latex}{abntex}{abntex2}{atigo científico}, 
		colorlinks=true,       		
    	linkcolor=blue,          	
    	citecolor=blue,        		
    	filecolor=magenta,      		
		urlcolor=blue,
		bookmarksdepth=4
}
\makeatother

\makeindex

\setlrmarginsandblock{3cm}{3cm}{*}
\setulmarginsandblock{3cm}{3cm}{*}
\checkandfixthelayout

\SingleSpacing

\begin{document}

\selectlanguage{brazil}

\frenchspacing

\maketitle

\begin{resumoumacoluna}
Escoamentos viscosos n\~ao s\~ao, em geral, discutidos de maneira detalhada em cursos de f\'isica geral e b\'asica. Isso se deve, em parte, ao fato da equa\c c\~ao de Navier-Stokes admitir solu\c c\~ao anal\'itica apenas para alguns poucos casos restritos, enquanto problemas mais sofisticados s\'o podem ser resolvidos por meio de m\'etodos num\'ericos. Neste trabalho, apresentamos um t\'unel de vento simulado, \textit{i.e.}, apresentamos um conjunto de programas capazes de resolver a equa\c c\~ao de Navier-Stokes para um objeto de formato arbitr\'ario inserido em um t\'unel de vento. O t\'unel nos possibilita visualizar a forma\c c\~ao de v\'ortices atrás do objeto, os conhecidos vórtices de von K\'arm\'an. É possível também calcular a for\c ca de arrasto sobre o objeto. Acreditamos que este t\'unel de vento num\'erico possa subsidiar o professor e permitir que uma discuss\~ao mais elaborada de escoamentos viscosos seja realizada. As potencialidades do t\'unel s\~ao exemplificadas por meio do estudo do comportamento da for\c ca de arrasto sobre um modelo simplificado de asa cujo \^angulo de ataque pode ser controlado. Um link para o download dos programas que comp\~oem o t\'unel \'e apresentado ao final do texto. 
 
 \vspace{\onelineskip}
 
 \noindent
 \textbf{Palavras-chave}: din\^amica dos fluidos, escoamentos viscosos, t\'uneis de vento, for\c ca de arrasto.
\end{resumoumacoluna}

\renewcommand{\resumoname}{Abstract}
\begin{resumoumacoluna}
Flow of viscous fluids are not usually discussed in detail in general and basic courses  of physics. This is due in part to the fact that the Navier-Stokes equation has analytical solution only for a few restricted cases, while more sophisticated problems can only be solved by numerical methods. In this text, we present a computer simulation of wind tunnel, \textit{i.e.}, we present a set of programs to solve the Navier-Stokes equation for an arbitrary object inserted in a wind tunnel. The tunnel enables us to visualize the formation of vortices behind object, the so-called von K\'arm\'an vortices, and calculate the drag force on the object. We believe that this numerical wind tunnel can support the teacher and allow a more elaborate discussion of viscous flow. The potential of the tunnel is exemplified by the study of the drag on a simplified model of wing whose angle of attack can be controlled. A link to download the programs that make up the tunnel appears at the end.
 \begin{otherlanguage*}{english}

   \vspace{\onelineskip}
 
 \end{otherlanguage*}  
\end{resumoumacoluna}


\textual

\section*{Introdução}
\addcontentsline{toc}{section}{Introdução}

Pode uma bolinha em queda frear? Um experimento recente \cite{Oliveira10} mostra-nos que, surpreendentemente, a resposta \'e sim. Os autores sugerem que isto se deve \`a forma\c c\~ao gradativa de uma esteira de v\' ortices no ar atr\'as da bolinha. Estes v\'ortices, que surgem em virtude da viscosidade do fluido, s\~ao arrastados pela bolinha em queda e, por isso, s\~ao respons\'aveis pela for\c ca de arrasto que se op\~oe ao movimento. No in\'icio do processo de forma\c c\~ao da esteira, a for\c ca de arrasto \'e proporcional \`a velocidade. Uma vez que a esteira tenha sido completada, a for\c ca se torna proporcional ao quadrado da velocidade. Na transi\c c\~ao entre um regime e outro, o arrasto torna-se maior que o peso e a frenagem \'e observada na bolinha. Este cen\'ario \'e confirmado por meio de simulações numéricas \cite{Oliveira12, Souza15}. 

Embora interessantes e prof\'icuos, veja por exemplo \cite{Aguiar04,Leroy77}, problemas f\'isicos que envolvem din\^amica de fluidos viscosos n\~ao s\~ao, em geral, discutidos em cursos de f\'isica b\'asica e geral. Quando isso ocorre, o assunto \'e tratado de maneira superficial \cite{Nussenzveig08,Schneiderbauer14}. Por quê? Para responder a esta pergunta precisamos primeiro definir o que são fluidos viscosos e discutir como são descritos.

Em mecânica dos fluidos, costuma-se definir como fluido qualquer substância que evidentemente não seja um sólido. Quando submetido a uma tensão de cisalhamento (tipo de tensão gerada por forças que atuam em sentidos opostos) um fluido tende a se deformar. A viscosidade, uma propriedade do fluido que determina quão facilmente o fluido se deforma, é uma função do fluido em questão e da temperatura e corresponde ao atrito interno resultante das interações entre as moléculas \cite{Schneiderbauer14, Munson90}.

Fluidos são descritos por meio de uma equação derivada independentemente por G. G. Stokes (1816 - 1903) e L. M. H. Navier (1785-1836), a equação de Navier-Stokes \cite{Feynman65}
%
\footnote{Pode-se encontrar facilmente na literatura deduções fenomenológicas da equação de Navier-Stokes, embora não haja uma dedução a partir de primeiros princípios aplicados à estrutura atômica microscópica do fluido. Veja, por exemplo, \cite{Schneiderbauer14,Feynman65}.
}:
\begin{equation}\label{navierstokes}
\frac{\partial \vec{\Omega}}{\partial t} = \frac{1}{\Rey} \nabla^2\vec{\Omega} - \vec{\nabla}\times(\vec{\Omega}\times\vec{v}),
\end{equation}  
em que $\vec{v}$ é o campo de velocidades, $\vec{\Omega} = \nabla \times \vec{v}$ \'e a vorticidade e $\Rey$ \'e o n\'umero de Reynolds, uma grandeza adimensional resultante da divis\~ao de for\c cas de in\'ercia por for\c cas de viscosidade em um elemento de fluido. Explicitamente,
\begin{equation}
\Rey = \frac{\rho v \textit{D}}{\mu},
\end{equation}
em que v é a velocidade do vento (longe do obstáculo, para o caso de um objeto inserido no interior de um túnel de vento), $\textit{D}$ é uma dimensão característica do objeto, $\mu$ é a viscosidade e $\rho$ é a massa específica do fluido. O número ou coeficiente de Reynolds foi introduzido por G. G. Stokes em 1851. Contudo, foi O. Reynolds (1842 - 1912) que popularizou o seu uso em 1883 e demonstrou, pela primeira vez, que o número poderia ser utilizado como um critério para distinção entre escoamento laminar, no qual as trajetórias das partículas do fluido tendem a ser paralelas, e turbulento, no qual as trajetórias das partículas são curvas, irregulares e entrecruzadas \cite{Munson90,Rott90,Reynolds83}. Por exemplo, para o escoamento sobre um cilindro, a transição entre o regime laminar e o turbulento ocorre quando $140 \leq \Rey \leq 300$ \cite{Blevins90,Humphreys60}.

A equação de Navier-Stokes admite soluções analíticas para alguns poucos casos. Na verdade, a prova matemática da existência de uma solução global para a equação de Navier-Stokes ainda não existe e é um dos \textit{Millennium Prize Problems} \cite{Schneiderbauer14}. Com isso, resta-nos recorrer às simulações numéricas. Embora m\'etodos num\'ericos para solu\c c\~ao da equa\c c\~ao de Navier-Stokes sejam numerosos na literatura (veja, por exemplo, \cite{Engelman90,Alfonsi02,Suh99,Pesavento04}), simula\c c\~oes voltadas para o ensino s\~ao bem menos abundantes. Neste texto apresentamos um t\'unel de vento simulado no qual se pode escolher o formato do objeto a ser inserido no t\'unel assim como a velocidade do vento. Imaginamos que esta ferramenta seja capaz de subsidiar o professor e permitir que uma discuss\~ao mais profunda de escoamentos viscosos seja realizada.

Este texto est\'a organizado da seguinte maneira. Inicialmente, os procedimentos para solu\c c\~ao da equa\c c\~ao de Navier-Stokes para um objeto de formato arbitr\'ario no interior de um t\'unel de vento s\~ao apresentados. Em seguida, descrevemos um m\'etodo, recentemente introduzido \cite{Souza15}, para determina\c c\~ao da for\c ca sobre objetos imersos em t\'uneis de vento. Posteriormente, exemplificamos potencialidades did\'aticas colocando no interior do túnel de vento uma asa cujo ângulo de ataque pode ser controlado. Finalmente, em uma se\c c\~ao concludente, apresentamos alguns coment\'arios que julgamos ser relevantes ao leitor e eventual utilizador do t\'unel simulado.

\section{Solu\c c\~ao da equa\c c\~ao de Navier-Stokes para um objeto de formato arbitr\'ario imerso em um t\'unel de vento}

O sistema f\'isico em quest\~ao \'e um t\'unel de vento, no qual a  velocidade do vento \'e control\'avel e onde pode ser	 inserido um objeto de formato arbitr\'ario, por exemplo, uma asa ou um cilindro. 

Para descrever a din\^amica deste sistema f\'isico, resolvemos a equação \ref{navierstokes} em uma rede bidimensional de 400 x 200 pontos (correspondente a uma região que denominamos \textit{região de interesse}), suficiente para nossos propósitos (a resolução poderá ser ampliada no caso de necessidade de maior precisão). Para resolver a equa\c c\~ao \ref{navierstokes}, usamos um m\'etodo \cite{Oliveira12} de diferen\c cas finitas com relaxa\c c\~oes sucessivas sobre as vorticidades
\footnote{Aplicamos o método das relaxações sucessivas aqui para resolver as equações de Stokes ($\Rey = 0 $) e de Navier-Stokes ($\Rey \neq 0 $). No primeiro caso, o método consiste simplesmente na substituição da vorticidade em um ponto da rede pela média das vorticidades dos seus quatro vizinhos imediatos. Para o caso da equação de Navier-Stokes, outros termos aparecem e estes também precisam ser expressos em função dos valores das velocidades e vorticidades nos pontos em seu entorno, tantos pontos quantos forem necessários para que todas as diferenças finitas que substituirão as derivadas tenham precisão pelo menos até a segunda ordem no passo discreto espacial e no passo de tempo. Para uma explana\c c\~ao do uso de relaxa\c c\~oes sucessivas na solu\c c\~ao de equa\c c\~oes diferenciais, veja \cite{Landau06,Pang06,Scherer05}. Curiosamente, na referência \cite[Problem~3.29]{Purcell65}, o método das relaxações é aplicado na solução de um problema de eletrodinâmica, o que traz à tona o conhecido fato de que a hidrodinâmica e a eletrodinâmica são formalmente muito semelhantes.}.
As configura\c c\~oes iniciais dos campos de velocidades e vorticidades s\~ao as resultantes da solu\c c\~ao da equa\c c\~ao \ref{navierstokes} no limite em que $\Rey \rightarrow 0$. Neste limite, a equa\c c\~ao de Navier-Stokes reduz-se a equa\c c\~ao de Laplace. Utilizar esta configura\c c\~ao inicial \'e necess\'ario pois do contr\'ario, um comportamento transiente esp\'urio pode ser obtido. Veja, por exemplo, \cite{Cruchaga02}.

As condições de contorno são: (a) na superfície do obstáculo e dentro dele, são mantidas nulas as velocidades e vorticidades; (b) fora da região de interesse (definida acima), o vento tem apenas uma componente não nula, ao longo da direção X, de modo que nesta região, $\vec v = V\hat x$ (por conveniência, fez-se $V = 1$ em unidades adimensionais) e as vorticidades são nulas.

\begin{figure}
	\centerline{\includegraphics[width=0.5\textwidth]{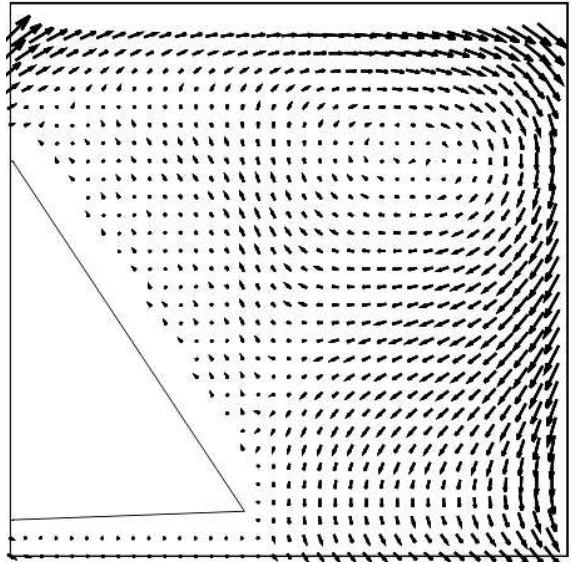}}
	\caption{V\'ortice aparecendo atr\'as de um objeto imerso no t\'unel.}
	\label{figura0}
\end{figure}

Em princ\'ipio, objetos de qualquer formato podem ser inseridos no t\'unel resguardada a condi\c c\~ao de que suas dimens\~oes n\~ao ultrapassem $20\%$ do tamanho total do
t\'unel, por exemplo, ou fração menor ainda caso seja necessário maior precisão. Ademais, o objeto precisa estar localizado no centro do t\'unel. Atendidas estas condi\c c\~oes, fomos bem sucedidos em observar a forma\c c\~ao da esteira de v\'ortices para $\Rey$ $\leq 1000$, sendo poss\'ivel simular com n\'umeros de Reynolds próximos de 1000 sem que altera\c c\~oes significativas no tamanho da rede, nas condi\c c\~oes de contorno e na viscosidade num\'erica sejam necess\'arias
\footnote{A introdu\c c\~ao da viscosidade num\'erica \'e um procedimento muito usado tradicionalmente para suavizar as varia\c c\~oes  
espaciais bruscas no campo de velocidades. Trata-se simplesmente de acrescentar \`as vorticidades uma pequena fra\c c\~ao de seu pr\'oprio Laplaciano. A viscosidade num\'erica foi utilizada por n\'os na solu\c c\~ao da equa\c c\~ao de Navier-Stokes para $\Rey>0$. Sua utiliza\c c\~ao produz uma suaviza\c c\~ao num\'erica, mas pode introduzir desvios quantitativos no resultado. Em princ\'ipio, a fra\c c\~ao do Laplaciano somada ao campo de vorticidades deve ser pequena o suficiente para que tais desvios estejam controlados. No caso de $\Rey=1000$, a fração adotada 
foi $1/100$.} .
A figura \ref{figura0} mostra um v\'ortice surgindo atr\'as de um objeto imerso no t\'unel, neste caso, uma asa. No link: \url{https://www.dropbox.com/s/1s4jbk589df14qp/tunel_asa_1000_22.wmv?dl=0} pode-se baixar um vídeo onde é mostrada a evolução do campo de velocidades quando, partindo da configuração de Stokes, o vento é ligado em $\Rey = 1000$ e escoa sobre uma asa com o ângulo de ataque acionado em $22º$. A evolução temporal observada atesta a adequação de nossos resultados a outros, experimentais e computacionais, presentes na literatura \cite{Feynman65,Blevins90,Alfonsi02,You07,Inoue99}.  

Contudo, como \'e poss\'ivel determinar a for\c ca de arrasto sobre o objeto no t\'unel? A pr\'oxima se\c c\~ao \'e dedicada \`a descri\c c\~ao de um m\'etodo geral capaz de faz\^e-lo \cite{Souza15}.

\section{Um m\'etodo alternativo para determina\c c\~ao da for\c ca sobre objetos imersos em t\' uneis de vento}

A maneira tradicional de calcular a for\c ca de arrasto sobre um obst\'aculo r\'igido envolve o conhecimento de dois elementos b\'asicos: o campo de press\~oes e a viscosidade nas proximidades da interface objeto/fluido, \textit{i.e.}, na regi\~ao correspondente \`a camada limite \cite{Landau87}. \'E poss\'ivel, alternativamente, obter a for\c ca de arrasto unicamente a partir do conhecimento do campo de velocidades do fluido em torno de um obst\'aculo, por meio do gradiente deste campo ao longo da superf\'icie da interface objeto/fluido. Neste caso, deve-se realizar uma integra\c c\~ao sobre esta superf\'icie \cite{Munson90,Noca99, Tan05}. Quando o campo de velocidades \'e determinado em pontos de uma rede discreta, a precis\~ao para o c\'alculo do gradiente fica comprometida, a menos que se adote uma rede muito fina perto da superfície de integra\c c\~ao, o que requer um grande esfor\c co computacional. Para contornar este problema t\'ecnico, propomos substituir a integral de superf\'icie por uma integral de volume sobre todo o volume ocupado pelo objeto, como descrevemos a seguir.

\begin{figure}
	\centerline{\includegraphics{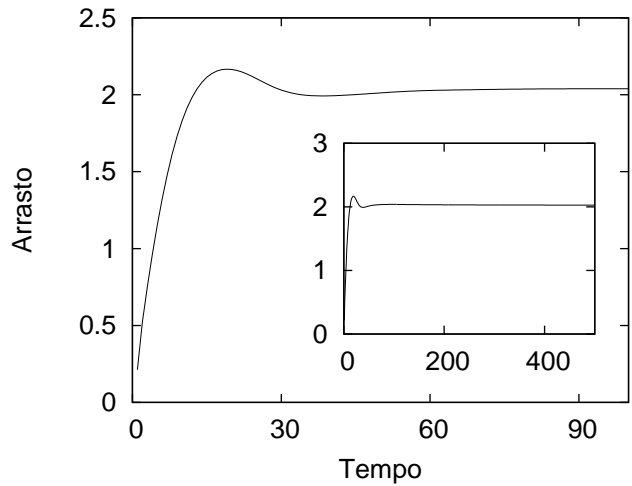}}
	\caption{For\c ca de arrasto sobre um cilindro est\'atico no t\'unel de vento. Inicialmente o t\'unel est\'a desligado. Em seguida, em $t = 0$, o t\'unel \'e ligado com n\'umero de  Reynolds $ \Rey = 1000 $. A componente da for\c ca na dire\c c\~ao perpendicular ao fluxo do vento \'e desprezível. A for\c ca de arrasto \'e representada em unidades arbitr\'arias, ignorando fatores como a densidade do fluido, etc. Durante uma unidade de tempo, o vento viaja o equivalente ao di\^ametro do cilindro. A componente da for\c ca que \'e paralela ao fluxo de vento aumenta, atinge um valor m\'aximo, diminui e finalmente se estabiliza, o que corresponde a uma situa\c c\~ao din\^amica em que aparecem continuamente v\'ortices sucessivos girando em sentidos alternados atr\'as do cilindro. A esteira de v\'ortices de von K\'arm\'an \'e ent\~ao formada. O destaque mostra o mesmo numa escala longa no tempo. }
	\label{figura1}
\end{figure}

No instante $t$, o campo de velocidades em cada ponto $\vec{r}$ da grade \'e $\vec{v}_t$$(\vec{r})$. 
Esta configura\c c\~ao no instante $t$ \'e obtida a partir de configura\c c\~oes em instantes anteriores $ t - \textit{d}t $, $t-2 \textit{d} t $, etc. Nos pontos da grade no interior do obst\'aculo  e na sua superfície, $\vec{v}_t = 0 $ em qualquer instante $t$. Suponhamos que o objeto seja removido no instante $t$ e o volume que o objeto ocupava seja preenchido com fluido est\'atico. Partindo da configura\c c\~ao do campo de velocidades j\'a conhecido $\vec{v}_t$$(\vec{r})$, obt\'em-se a configura\c c\~ao futura no tempo $t + \textit{D}t$ . Como o objeto r\'igido foi substitu\'ido por um fluido, algumas velocidades n\~ao nulas aparecem no interior do volume anteriormente ocupado pelo objeto. Em outras palavras, a remo\c c\~ao do objeto permite que o campo de velocidades penetre um pouco no interior do seu volume. Agora, este campo de velocidades interno pode ser integrado no volume. O resultado \'e multiplicado pela densidade do fluido, obtendo-se assim o momento que seria transferido do fluido para o objeto. Dividindo-se esse impulso por $\textit{D}t$, finalmente obt\'em-se a for\c ca de arrasto 
\footnote {O tempo $\textit{D}t$ necess\'ario para que o vento entre na regi\~ao anteriormente ocupada pelo objeto varia de acordo com o n\'umero de Reynolds e com as condi\c c\~oes iniciais. Considera-se que uma estimativa do tempo \'e aceit\'avel, quando multiplicado por dois, o momento transferido torna-se duas vezes maior.}. 
Pode-se observar na figura \ref {figura1} a for\c ca de arrasto sobre um longo cilindro
est\'atico perpendicular ao vento no interior do t\'unel, em fun\c c\~ao do tempo; esta for\c ca corresponde aproximadamente \`a situa\c c\~ao f\'isica que descrevemos na se\c c\~ao introdut\'oria deste texto. O comportamento da força de arrasto para o cilindro estático se harmoniza com a interpretação sugerida na seção introdutória para o problema da bolinha em queda: há um tempo transiente durante o qual a esteira de von Kármán ainda não se formou e a força de arrasto não é proporcional a $v^2$, o que ocorre tão logo a esteira esteja formada. Na transição entre estes regimes, a força de arrasto torna-se maior que o peso da bolinha e a mesma experimenta uma frenagem. 

Naturalmente, o m\'etodo acima descrito pode ser aplicado n\~ao apenas a um cilindro, mas a um objeto de formato qualquer. 
\section{Uma asa em um t\'unel de vento}

Para exemplificar as potencialidades do t\'unel, apresentamos, nesta se\c c\~ao, um estudo do comportamento da for\c ca de arrasto sobre um modelo simplificado de asa cujo \^angulo de ataque, \textit{i.e.}, o \^angulo que o eixo da asa faz com o vento, pode ser controlado.

\begin{figure}
\centerline{\includegraphics[width=0.5\textwidth]{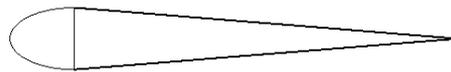}}
	\caption{Modelo simplificado de asa introduzida no t\'unel. O \^angulo de ataque pode ser controlado entre $\pm 22º$.}
	\label{figura2}
\end{figure}

Nossa simula\c c\~ao corresponde \`a situa\c c\~ao f\'isica de um aeromodelo, cuja asa tem largura caracter\'istica aproximada de $60 cm$  voando a uma velocidade aproximada de $24  m/s$ no ar, cuja viscosidade \'e de  $1.4207.10^{-5} m^2/s$. A asa que desenhamos \'e mostrada na figura \ref{figura2}.
Para a velocidade do vento fixa em $\Rey=1000$, o \^angulo de ataque pode variar entre $\pm 22º$ sem que as paredes do túnel interfiram fortemente na simula\c c\~ao. Neste caso, ap\'os o regime transiente, v\'ortices girando em sentidos contr\'arios aparecem alternadamente atr\'as da asa, assim como ocorre com o cilindro.

\begin{figure}
\begin{center}
\includegraphics[width=0.5\textwidth]{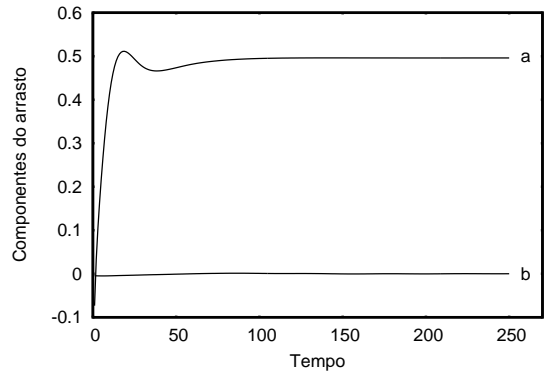}
	\caption{Componentes do arrasto para \^angulo de ataque igual a $0º$. Em (a),  apresenta-se a componente longitudinal e, em (b), a componente transversal. A componente longitudinal cresce, atinge um valor m\'aximo, reduz-se um pouco e se estabiliza, passando a oscilar levemente (impercept\'ivel na figura) em torno de um valor fixo positivo, enquanto a componente transversal, após o transiente, passa a flutuar em torno de um valor fixo, no caso do ângulo de ataque igual a $0º$, este valor é nulo, como esperado para obstáculos simétricos.}
	\label{figura3}
	\end{center}
\end{figure}

\begin{figure}
\begin{center}
\includegraphics[width=0.5\textwidth]{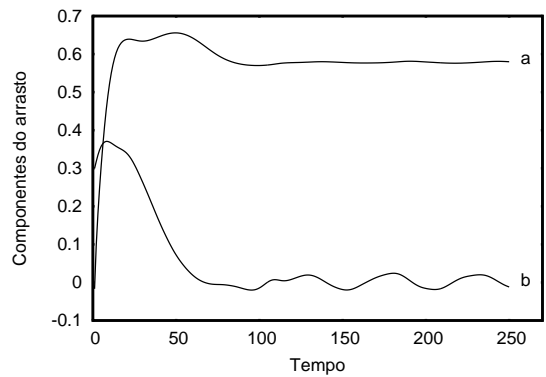}
	\caption{Componentes do arrasto para \^angulo de ataque igual a $15º$. Em (a),  apresenta-se a componente longitudinal e, em (b), a componente transversal. Assim como antes, a componente longitudinal cresce, atinge um valor m\'aximo, reduz-se um pouco e estabiliza, passando a oscilar em torno de um valor fixo positivo enquanto a componente transversal, após o transiente (neste caso, bem diferente do caso para \^angulo de ataque 0º, pois a simetria axial foi quebrada), passa a flutuar em torno de um valor fixo, no caso do ângulo de ataque igual a $15º$, este valor é positivo, embora pequeno se comparado com a componente longitudinal.}
	\label{figura4}
	\end{center}
\end{figure}

A componente longitudinal do arrasto neste caso apresenta um comportamento qualitativo an\'alogo ao observado para o cilindro (veja figura \ref{figura1}): a for\c ca cresce, atinge um valor m\'aximo, reduz-se um pouco e estabiliza, passando a oscilar em torno de um valor fixo positivo. Este valor fixo aumenta com o \^angulo de ataque. O comportamento da componente transversal do arrasto, igualmente, depende do \^angulo de ataque. Para um \^angulo de ataque igual a $0º$, observa-se que esta componente, depois de um est\'agio transiente, flutua continuamente em torno de zero, o que é esperado para um obstáculo axialmente simétrico \cite{Munson90}. \`A medida que o \^angulo de ataque cresce, os valores desta componente aumentam. Este resultado corresponde satisfatoriamente ao que se sabe sobre a rela\c c\~ao entre a sustenta\c c\~ao, for\c ca correspondente a componente transversal do arrasto, e o \^angulo de ataque \cite{Babinsky03,Weltner01}. Veja as figuras \ref{figura3} e \ref{figura4}.


\begin{figure}
\begin{center}
	\centerline{\includegraphics[width=0.5\textwidth]{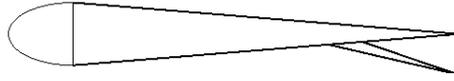}}
	\caption{Modelo simplificado de asa com flap.}
	\label{figura5}
	\end{center}
\end{figure}

\begin{figure}
\begin{center}
	\includegraphics[width=0.5\textwidth]{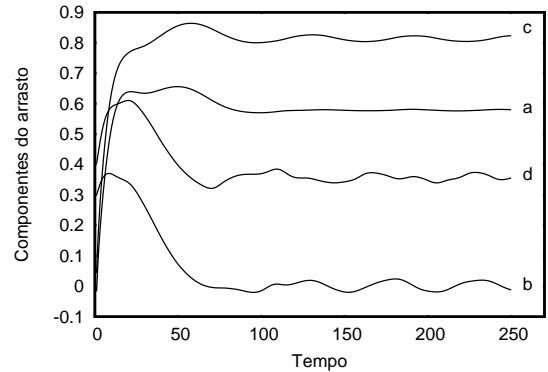}
	\caption{Comparação das componentes do arrasto para a asa com e sem flap para um \^angulo de ataque igual $15º$. Em (a) e (b), se observam, respectivamente, as componentes longitudinal e transversal do arrasto para asa \textbf{sem} flap (repetido da figura 5) enquanto em (c) e (d), se observam, respectivamente, as componentes longitudinal e transversal do arrasto para asa \textbf{com} flap. Como esperado, para um mesmo \^angulo de ataque, a asa com flap apresenta valores mais altos para suas componentes embora o comportamento qualitativo seja aproximadamente o mesmo com ou sem flap.}
	\label{figura6}
	\end{center}
\end{figure}

Em nossa simula\c c\~ao, a asa pode ser ligeiramente modificada para incluir um tipo de flap. Flaps consistem em abas ou superf\'icies articuladas existentes na parte posterior das asas que alteram temporariamente a geometria das mesmas. Quando abaixados e/ou estendidos, os flaps aumentam o arrasto e a sustenta\c c\~ao, o que \'e especialmente importante no procedimento de pouso. O flap que instalamos na asa \'e mostrado na figura \ref{figura5} e \'e equivalente a um flap do tipo ventral. As altera\c c\~oes nas componentes do arrasto em fun\c c\~ao da presen\c ca do flap para um \^angulo de ataque igual a $15º$ s\~ao mostradas na figura \ref{figura6}. Da an\'alise do gr\'afico, percebe-se que a presen\c ca do flap altera a for\c ca de arrasto, aumentando o valor de ambas as componentes em rela\c c\~ao ao seu valor sem o flap, o que se harmoniza com o que se sabe a respeito da fun\c c\~ao deste tipo de flap nas asas \cite{Munson90}.
%
\section*{Considerações finais}

O estudo de escoamentos viscosos, embora riqu\'issimo, \'e praticamente ausente em cursos de f\'isica geral e b\'asica. Isso se deve, em parte, ao fato deste tipo de escoamento ser descrito pela equa\c c\~ao de Navier-Stokes que, por sua vez, só admite solu\c c\~ao anal\'itica para alguns poucos casos. Diante deste quadro, apresentamos neste texto um t\'unel de vento simulado no qual se pode escolher o formato do objeto a ser inserido no t\'unel assim como a velocidade do vento. O objetivo do túnel é instrumentalizar o professor e viabilizar que uma discussão mais aprofundada do tema seja realizada. 

Os programas que comp\~oem a simula\c c\~ao, assim como um arquivo com instru\c c\~oes para utiliza\c c\~ao do t\'unel podem ser baixados no endere\c co \url{https://www.dropbox.com/s/t3y8og20zfhypk8/programas.rar?dl=0}. 

Alguns coment\'arios finais se fazem necess\'arios. O t\'unel, concebido para analisar escoamentos bidimensionais, pode ser facilmente modificado para dar conta de escoamentos tridimensionais. Por\'em, nossa experi\^encia com a utiliza\c c\~ao do t\'unel indica que isso exigiria enorme esfor\c co computacional, o que at\'e ent\~ao n\~ao pode ser contornado nem mesmo com a utiliza\c c\~ao de clusters de alto desempenho. Isso se deve ao fato de ser difícil  realizar um processo de computa\c c\~ao em paralelo para o t\'unel uma vez que uma configura\c c\~ao \'e gerada a partir da anterior.

No túnel, ``as paredes'' que limitam a região de interesse, inevitavelmente, interferem nos resultados. Para minimizar sua influência, seria necessário adotar paredes muito mais afastadas do objeto. Na verdade, para $\Rey$ pequenos, teríamos adicionalmente que adotar um túnel tridimensional, porque em duas dimensões, a influência das paredes se faz sentir até distâncias proporcionais a $1/\Rey$ \cite{Childress09, Shaw09}. Esta interferência, contudo, não modifica os principais aspectos qualitativos da solução, aspectos estes que são o foco deste trabalho.

Num formato piloto, o t\'unel foi utilizado em duas disciplinas, f\'isica t\'ermica para o curso t\'ecnico de automa\c c\~ao industrial e f\'isica geral II para licenciatura em f\'isica do Instituto Federal do Rio de Janeiro. Na aplica\c c\~ao no curso de automa\c c\~ao, por tratar-se de um curso de n\'ivel m\'edio, o t\'opico escoamentos viscosos foi apresentado como complementar. Neste caso, o problema da bolinha que freia ao cair foi apresentado e discutido em detalhes com o aux\'ilio do t\'unel. Simula\c c\~oes com asas tamb\'em foram utilizadas para discutir as raz\~oes f\'isicas pelas quais o avi\~ao consegue manter-se no ar. No curso de licenciatura, o t\'unel foi utilizado por um pequeno n\'umero de alunos para elabora\c c\~ao do projeto de final do curso. Neste caso, inicialmente, os alunos se familiarizaram com t\'unel num\'erico. Posteriormente, inseriram diferentes objetos no t\'unel, como uma bala de revolver, um tri\^angulo e um cilindro girando com velocidade angular control\'avel. Os dados obtidos por eles para este \'ultimo caso foram contrapostos com os presentes em \cite{Souza15}, o que viabilizou uma interessante discuss\~ao sobre o efeito Magnus e suas manifesta\c c\~oes.  

\section{Agradecimentos}
Os autores agradecem a P. M. C. Dias e ao primeiro árbitro pelas críticas e sugestões ao manuscrito. Os autores também agradecem as agências de fomento CAPES e CNPQ pelo financiamento.

\bibliography{abntex2-modelo-references}

\end{document}